\documentclass[a4paper]{jpconf}
\usepackage{graphicx}

\usepackage{amsmath} 
\usepackage{amssymb} 
\usepackage{bm} 

\begin{document}
\title{Excitation and transport of bound magnon clusters in frustrated ferromagnetic chain}

\author{Hiroaki Onishi}

\address{Advanced Science Research Center, Japan Atomic Energy Agency, Tokai, Ibaraki 319-1195, Japan}

\ead{onishi.hiroaki@jaea.go.jp}

\begin{abstract}
We investigate magnetic excitations and spin transport properties of a frustrated spin chain
with ferromagnetic nearest-neighbor and antiferromagnetic next-nearest-neighbor exchange couplings
in a magnetic field
by density-matrix renormalization group methods.
In the field-induced quadrupole and octupole regimes,
the low-energy excitation is governed by two- and three-magnon bound states, respectively,
so that bound magnon clusters would contribute to the spin transport.
We show that spin current correlations decrease
when the system goes from the quadrupole regime into the octupole regime.
This indicates that the spin transport is suppressed in the octupole regime,
although bound three-magnon clusters carry a larger amount of angular momentum
than bound two-magnon clusters.
\end{abstract}

\section{Introduction}

Spin-related transport phenomena in magnetic materials have been attracting a growing interest
in the fields of condensed-matter physics and spintronics.
In magnetic insulators,
magnons,
which are elementary excitations in magnetic ordered states,
carry spin current without electric current,
expected to be a key feature to realize efficient devices without Joule heating.
Recently,
novel types of carriers in quantum disordered states have been discussed,
such as
spinons~\cite{Hirobe2016,Bertini2021},
bound magnon pairs~\cite{Onishi2019,Hirobe2019},
and Majorana fermions~\cite{Koga2020,Minakawa2020},
not only to understand fundamental properties of exotic spin states from the viewpoint of transport
but also to explore a new mechanism for the efficient spin current generation.

Here we focus on spin multipole states emerging in a frustrated spin chain
with ferromagnetic nearest-neighbor and antiferromagnetic next-nearest-neighbor exchange couplings
in a magnetic field
\cite{Kecke2007,Hikihara2008,Sudan2009}.
It is well established that
the ground state is a vector chiral state in low magnetic fields,
while in high magnetic fields
there appear a series of spin multipole liquid ground states
such as quadrupole, octupole, and hexadecapole,
which are two-, three-, and four-magnon bound states, respectively.
Since the low-energy excitation is governed by multi-magnon bound states,
we expect that bound multi-magnon clusters would carry spin current.

We have previously studied
the dynamical spin and quadrupole structure factors
\cite{Onishi2015a,Onishi2015b,Onishi2018}
and the spin Drude weight
\cite{Onishi2019}
with a particular focus on the quadrupole regime.
In this paper,
to gain further insight into the relation between the low-energy excitation and the spin transport property,
we numerically investigate
the multi-magnon excitation energy,
the dynamical spin, quadrupole, and octupole structure factors,
and the spin current correlation function,
in the both quadrupole and octupole regimes.
We show that spin current correlations are suppressed in the octupole regime,
although bound three-magnon clusters carry a larger amount of angular momentum
than bound two-magnon clusters.

\section{Model and method}

We consider a spin-1/2 $J_{1}$-$J_{2}$ Heisenberg model
with competing ferromagnetic $J_{1}$ ($<$0) and antiferromagnetic $J_{2}$ ($>$0)
in a magnetic field $h$
on an $N$-site chain,
described by
\begin{equation}
  H =
  J_{1} \sum_{i} \bm{S}_{i} \cdot \bm{S}_{i+1}
  + J_{2} \sum_{i} \bm{S}_{i} \cdot \bm{S}_{i+2}
  - h \sum_{i} S_{i}^{z},
\label{eq: H}
\end{equation}
where $\bm{S}_{i}$ are spin-1/2 operators at site $i$.
We set $J_{2}=1$ and take it as the energy unit.
Note that the total magnetization $m=M/N=\sum_{i}S_{i}^{z}/N$ is conserved,
so that it can be used to block-diagonalize the Hamiltonian.
In the magnetization curve,
$M$ changes by one with $h$ in the vector chiral phase,
two in the quadrupole liquid phase,
three in the octupole liquid phase, etc.
For the analysis of a given $m$,
we set $h$ to the midpoint of the magnetization plateau of $M$
in the $N$-site system,
and use the saturation field $h_{\mathrm{s}}$ for $m=1/2$.

We investigate magnetic excitations and spin transport properties at zero temperature
by density-matrix renormalization group (DMRG) methods
in the open boundary condition
\cite{White1992,Jeckelmann2002}.
To clarify the property of the low-energy excitation,
we examine the dependence of the $p$-magnon excitation energy, defined by
\begin{equation}
  \Delta_{p}(N,M) = [E_{0}(N,M+p)+E_{0}(N,M-p)-2E_{0}(N,M)]/2,
\end{equation}
where $E_{0}(N,M)$ is the lowest energy
of the $N$-site system in the subspace of $M$ at $h=0$.
Note that excitation energies for adding $p$ magnons and removing $p$ magnons are averaged.
At the saturation $M=N/2$, we evaluate the $p$-magnon excitation energy as
\begin{equation}
  \Delta_{p}(N,N/2) = E_{0}(N,N/2-p)-E_{0}(N,N/2)+ph_{\mathrm{s}}
\end{equation}
Moreover, we calculate the dynamical spin, quadrupole, and octupole structure factors,
which are, respectively, given by
\begin{equation}
  S^{-}(q,\omega) =
  -\mathrm{Im}\frac{1}{\pi}
  \langle \psi_{\mathrm{G}} \vert
  (S_{q}^{-})^{\dag}
  \frac{1}{\omega+E_{\mathrm{G}}-H+{\rm i}\eta}
  S_{q}^{-}
  \vert \psi_{\mathrm{G}} \rangle,
\end{equation}
\begin{equation}
  Q^{--}(q,\omega) =
  -\mathrm{Im}\frac{1}{\pi}
  \langle \psi_{\mathrm{G}} \vert
  (Q_{q}^{--})^{\dag}
  \frac{1}{\omega+E_{\mathrm{G}}-H+{\rm i}\eta}
  Q_{q}^{--}\
  \vert \psi_{\mathrm{G}} \rangle,
\end{equation}
\begin{equation}
  O^{---}(q,\omega) =
  -\mathrm{Im}\frac{1}{\pi}
  \langle \psi_{\mathrm{G}} \vert
  (O_{q}^{---})^{\dag}
  \frac{1}{\omega+E_{\mathrm{G}}-H+{\rm i}\eta}
  O_{q}^{---}
  \vert \psi_{\mathrm{G}} \rangle,
\end{equation}
where $S_{q}^{-}$, $Q_{q}^{--}$, and $O_{q}^{---}$ are
the Fourier transforms of
$S_{i}^{-}$, $Q_{i}^{--}=S_{i}^{-}S_{i+1}^{-}$, and $O_{i}^{---}=S_{i}^{-}S_{i+1}^{-}S_{i+2}^{-}$,
respectively,
$\vert \psi_{\mathrm{G}} \rangle$ is the ground state,
$E_{\mathrm{G}}$ is the ground-state energy,
and $\eta$ is a small broadening factor,
set to 0.1 in the present calculations.
To calculate the dynamical structure factors,
we use a dynamical DMRG method
\cite{Jeckelmann2002},
multi-targeting
$\vert \psi_{\mathrm{G}} \rangle$,
$A \vert \psi_{\mathrm{G}} \rangle$,
and
$[\omega+E_{\mathrm{G}}-H+\mathrm{i}\eta]^{-1} A \vert \psi_{\mathrm{G}} \rangle$,
where
$A$ is $S_{q}^{-}$, $Q_{q}^{--}$, or $O_{q}^{---}$.

Regarding the transport property,
we investigate the spin current correlation function
as a measure of the degree of the flow of the spin current,
defined by
\begin{equation}
  I_{\mathrm{s}}=
  \langle \psi_{\mathrm{G}} \vert
  j_{\mathrm{s}}^{\dag} j_{\mathrm{s}}
  \vert \psi_{\mathrm{G}} \rangle,
\end{equation}
where
$j_{\mathrm{s}}
=\sum_{i} J_{1} (\bm{S}_{i}\times\bm{S}_{i+1})^{z}
+\sum_{i} 2J_{2} (\bm{S}_{i}\times\bm{S}_{i+2})^{z}$
is the spin current operator.

\section{Results}

\begin{figure}[t]
\begin{center}
\includegraphics[scale=0.7]{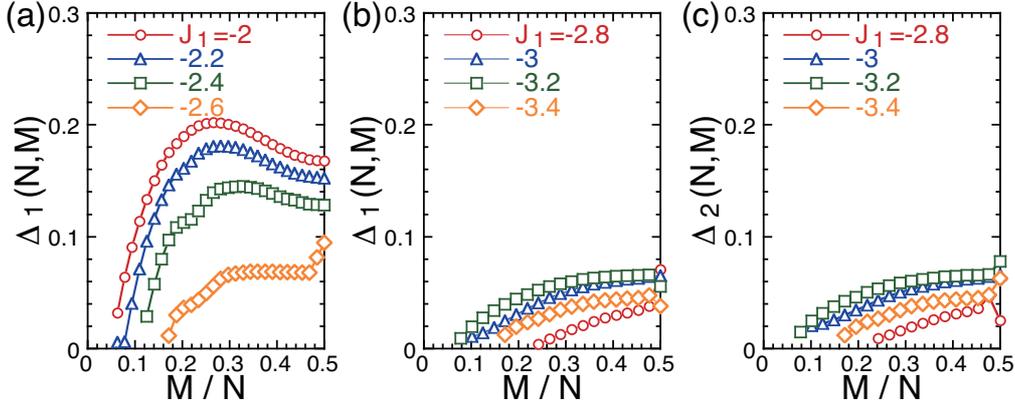}
\end{center}
\caption{
The excitation energy
(a) $\Delta_{1}(N,M)$
for several values of $J_{1}$ in the quadrupole regime,
(b) $\Delta_{1}(N,M)$ and (c) $\Delta_{2}(N,M)$
for several values of $J_{1}$ in the octupole regime.
The system size is $N=128$.
}
\label{fig1}
\end{figure}

Let us start with the discussion on magnetic excitations.
In Fig.~\ref{fig1}(a),
we show the excitation energy $\Delta_{1}(N,M)$
for several values of $J_{1}$ in the quadrupole regime.
Note that the two-magnon bound state appears for $J_{1} \gtrsim -2.7$
and the ground state turns to the three-magnon bound state for smaller $J_{1}$
\cite{Kecke2007,Hikihara2008,Sudan2009}.
We find that $\Delta_{1}(N,M)$ decreases with decreasing $J_{1}$,
indicating that the two-magnon bound state becomes unstable
as the system approaches the phase boundary.
We note that $\Delta_{1}(N,M)$ increases with decreasing $J_{1}$ for larger $J_{1}$ $(>-2)$,
indicating a reentrant behavior
\cite{Onishi2015a}.

Figures~\ref{fig1}(b) and \ref{fig1}(c) present
the excitation energies $\Delta_{1}(N,M)$ and $\Delta_{2}(N,M)$, respectively,
in the octupole regime.
We find that $\Delta_{1}(N,M)$ and $\Delta_{2}(N,M)$ exhibit similar dependences.
This is because we obtain both single-spin-flip and two-spin-flip states
when we break a bound three-magnon cluster.
With decreasing $J_{1}$,
$\Delta_{1}(N,M)$ and $\Delta_{2}(N,M)$ first increase for $J_{1}=-2.8$, $-3$, and $-3.2$,
indicating that the ferromagnetic exchange interaction stabilizes the three-magnon bound state,
while they turn to decrease with further decreasing $J_{1}$.
This reentrant behavior resembles that of $\Delta_{1}(N,M)$ in the quadrupole regime.
Here we point out that the excitation energies in the octupole regime are much reduced
from those in the quadrupole regime.
This is indicative that the three-magnon bound state is easily broken
comparing with the two-magnon bound state.

\begin{figure}[t]
\begin{center}
\includegraphics[scale=0.7]{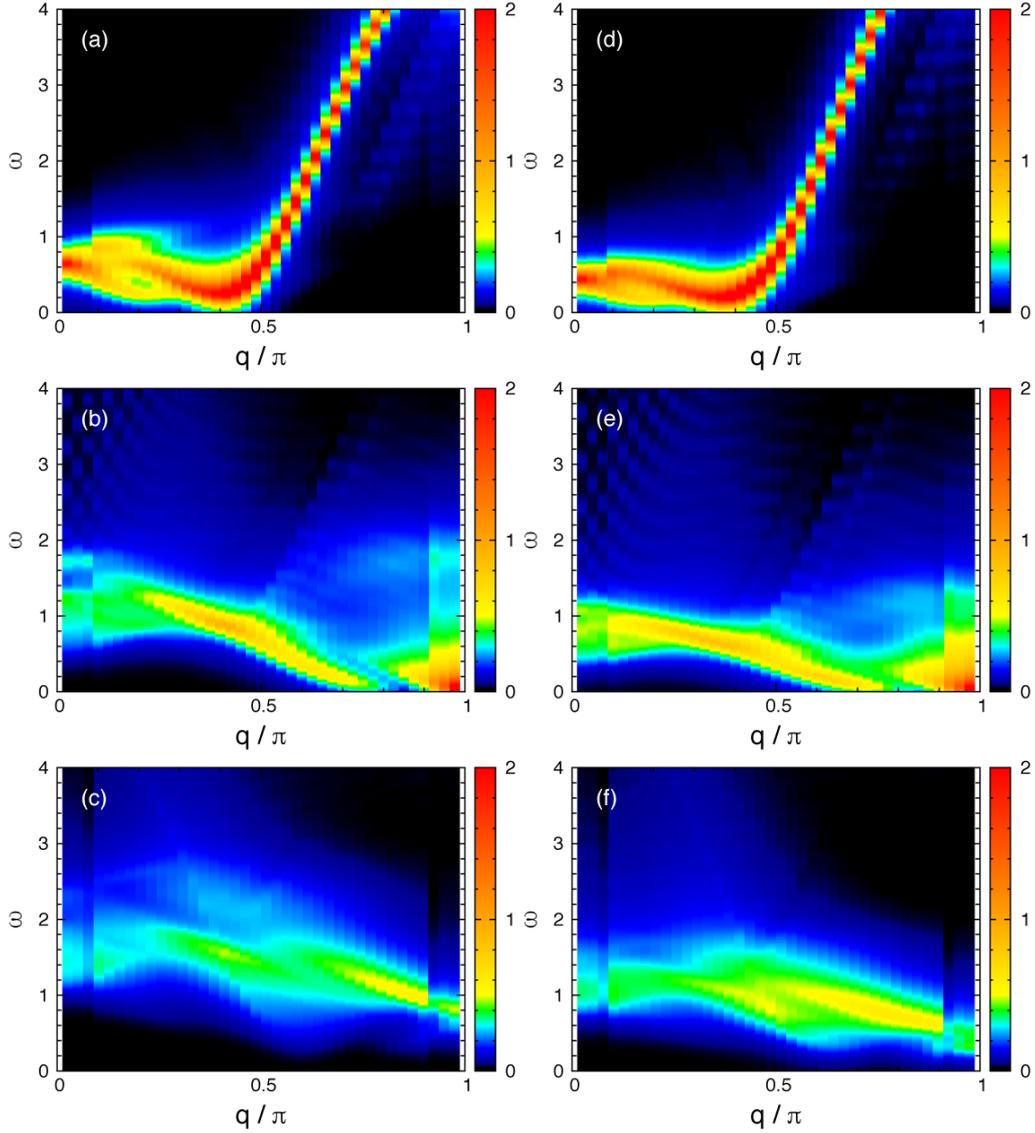}
\end{center}
\caption{
Intensity plots of the dynamical structure factors
for two values of $J_{1}$ with fixed $m=0.35$ in the quadrupole regime.
Left panels:
(a) $S^{-}(q,\omega)$,
(b) $Q^{--}(q,\omega)$,
and
(c) $O^{---}(q,\omega)$
for $J_{1}=-2$.
Right panels:
(d) $S^{-}(q,\omega)$,
(e) $Q^{--}(q,\omega)$,
and
(f) $O^{---}(q,\omega)$
for $J_{1}=-2.4$.
The system size is $N=40$.
}
\label{fig2}
\end{figure}

\begin{figure}[t]
\begin{center}
\includegraphics[scale=0.7]{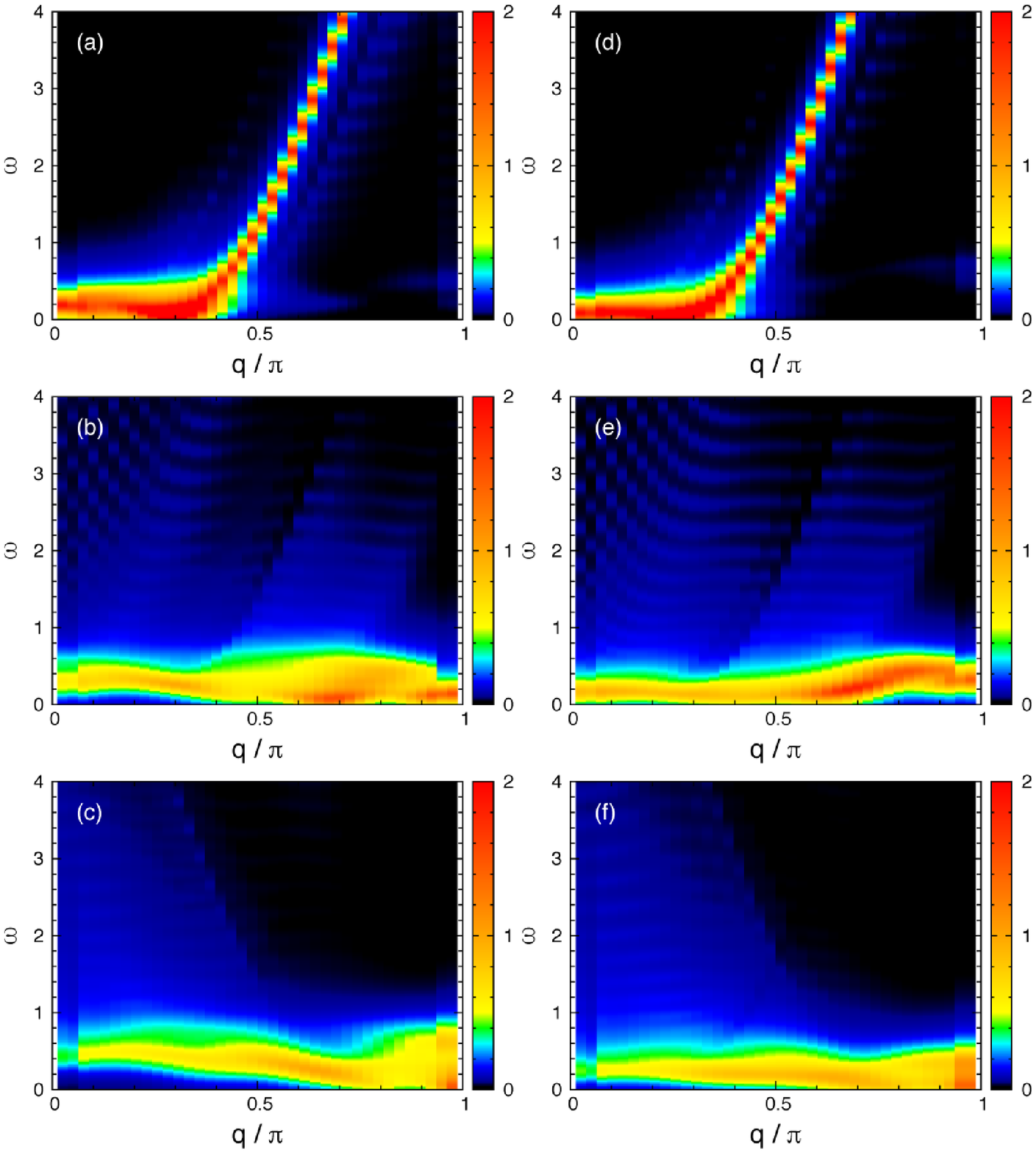}
\end{center}
\caption{
Intensity plots of the dynamical structure factors
for two values of $J_{1}$ with fixed $m=0.35$ in the octupole regime.
Left panels:
(a) $S^{-}(q,\omega)$,
(b) $Q^{--}(q,\omega)$,
and
(c) $O^{---}(q,\omega)$
for $J_{1}=-3$.
Right panels:
(d) $S^{-}(q,\omega)$,
(e) $Q^{--}(q,\omega)$,
and
(f) $O^{---}(q,\omega)$
for $J_{1}=-3.4$.
The system size is $N=40$.
}
\label{fig3}
\end{figure}

To obtain more deep insight into the low-energy excitation,
we investigate the dynamical spin, quadrupole, and octupole structure factors.
In Fig.~\ref{fig2},
we show numerical results for $J_{1}=-2$ and $-2.4$ with fixed $m=0.35$,
where the ground state is in the quadrupole regime.
For $S^{-}(q,\omega)$,
we find gapped spin excitations,
composed of a dispersive mode extending to high energy.
The momentum position of the gap,
corresponding to the bottom of the dispersion at a little below $q=\pi/2$,
is incommensurate.
With decreasing $J_{1}$,
the momentum position of the gap gradually moves toward small momentum,
while the gap decreases,
as observed in Fig.~\ref{fig1}(a).
In addition, the dispersion for small momentum below the gap position
is pushed down to the low-energy side.
For $Q^{--}(q,\omega)$,
we find gapless quadrupole excitations at $q=\pi$
due to quasi-long-range antiferro-quadrupole correlations.
The overall structure of the intensity distribution does not change with $J_{1}$ significantly,
while it moves to the low-energy side,
so that the slope of the dispersion becomes gentle.
For $O^{---}(q,\omega)$,
we find gapped octupole excitations.
The intensity distribution moves to the low-energy side
with decreasing $J_{1}$,
which is similar to the findings for $S^{-}(q,\omega)$ and $Q^{--}(q,\omega)$.

Figure~\ref{fig3} presents the dynamical structure factors
for $J_{1}=-3$ and $-3.4$ with fixed $m=0.35$,
where the ground state is in the octupole regime.
For $S^{-}(q,\omega)$,
we find gapped spin excitations.
Note that the lowest-energy peak is found at a finite energy,
but the gap is not clearly identified in the color plot
due to large intensity near zero energy.
With decreasing $J_{1}$,
the dispersion in small momentum goes further down to the low-energy side,
leading to a flat structure of the dispersion.
For $Q^{--}(q,\omega)$,
we find gapped quadrupole excitations.
The momentum position of the gap is away from $q=\pi$.
That is, dominant quadrupole correlations change from
quasi-long-range antiferro-quadrupole correlations in the quadrupole liquid
to short-range incommensurate quadrupole correlations in the octupole liquid.
The intensity is concentrated in a low-energy dispersion,
in contrast to the highly dispersive mode of $S^{-}(q,\omega)$.
For $O^{---}(q,\omega)$,
we find gapless octupole excitations at $q=\pi$
due to quasi-long-range antiferro-octupole correlations.
The intensity is concentrated in a low-energy dispersion
in similar to $Q^{--}(q,\omega)$.
Thus the slope of the low-energy dispersion becomes gentler
than that in the quadrupole regime.

\begin{figure}[t]
\begin{center}
\includegraphics[scale=0.6]{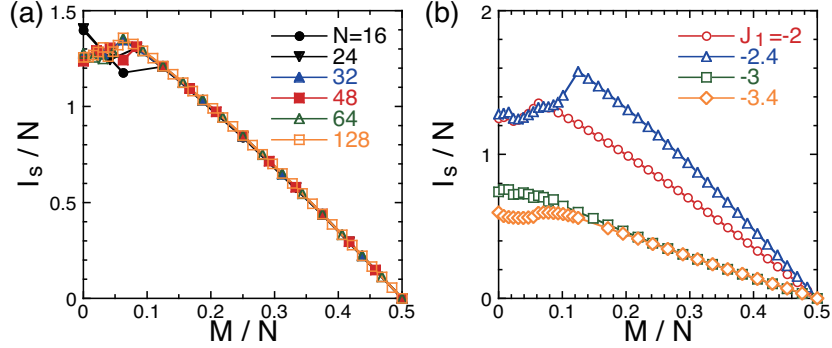}
\end{center}
\caption{
The spin current correlation function $I_{\mathrm{s}}/N$
(a) at $J_{1}=-2$ with various system sizes,
and
(b) the dependence on $J_{1}$ with $N=128$.
}
\label{fig4}
\end{figure}

Now we move on to the transport property.
In Fig.~\ref{fig4}(a),
we show typical results of the spin current correlation function
$I_{\mathrm{s}}/N$.
Plots for various system sizes collapse,
indicating that the finite-size effect is small.
In Fig.~\ref{fig4}(b),
we present $I_{\mathrm{s}}/N$ at $J_{1}=-2$ and $-2.4$ for the quadrupole regime,
and at $J_{1}=-3$ and $-3.4$ for the octupole regime.
For the quadrupole regime,
we observe that
the spin current correlation is enhanced with decreasing $J_{1}$,
implying that the system changes to be favorable for the flow of bound two-magnon clusters.
With further decreasing $J_{1}$,
we find that the spin current correlation in the octupole regime
is much reduced from that in the quadrupole regime,
indicating that the spin transport is suppressed.
This is contrary to a naive expectation,
because bound three-magnon clusters carry a larger amount of angular momentum
than bound two-magnon clusters.
This behavior would be attributed to the dependences of the low-energy dispersion.
That is,
the slope of the dispersion, corresponding to the velocity of propagating magnon clusters,
in the octupole regime is gentler than that in the quadrupole regime,
leading to the reduction of the flow of the spin current.

\section{Summary}

We have studied magnetic and transport properties
of the frustrated ferromagnetic chain by numerical methods.
The spin current correlation in the octupole regime is reduced from that in the quadrupole regime,
although bound three-magnon clusters carry a larger amount of angular momentum
than bound two-magnon clusters.

\section*{Acknowledgments}

This work was supported by JSPS KAKENHI Grant Number JP19K03678.
Computations were done using supercomputers at
the Japan Atomic Energy Agency and
the Institute for Solid State Physics, the University of Tokyo.

\section*{References}

\end{document}